# From Optical Breakdown to Bubble Inception: A Coupled Plasma-Thermal Framework for Nanosecond Laser-Induced Cavitation in Water

Shuqi Zhou, Abdol Hadi Mokarizadeh, Ben Xu*

*Department of Mechanical and Aerospace Engineering, University of Houston, Houston, Texas 77204, USA*

## ABSTRACT

Laser-induced cavitation under nanosecond optical breakdown is central to applications such as laser-induced forward transfer (LIFT), microsurgery, and microfluidic actuation, yet the physical origin of the earliest cavity and its connection to subsequent bubble growth remain unresolved. Existing models typically describe bubble formation either as a plasma-driven mechanical response or as a thermally driven nucleation process, without resolving how these mechanisms interact during inception. In this study, we developed a coupled plasma-thermal framework that unifies free-electron dynamics, plasma absorption, thermoelastic acoustic response, residual thermal energy retention, and post-inception bubble evolution within a single description. The model shows that bubble inception is governed primarily by plasma-induced thermoelastic acoustic relaxation, which generates transient tensile rarefaction pressures sufficient for cavitation on nanosecond timescales, while residual thermal energy sustains subsequent bubble growth. Because plasma-mediated energy deposition is spatially anisotropic under moving breakdown conditions, the initial cavity is predicted to form as an elongated, asymmetric bubble rather than as a spherical nucleus. This initial bubble shape reflects the spatial structure of the plasma absorption region and provides the starting geometry for subsequent bubble dynamics. Comparison with time-resolved experiments demonstrates that the coupled framework captures both early-time cavity formation and long-time bubble expansion more accurately than plasma-only or thermal-only models. These results show how breakdown-induced asymmetry controls the birth shape of the cavitation bubble and its early fluid-mechanical evolution, providing a physics-based basis for modeling laser-driven liquid motion and material transport.

**Keywords:** Laser-induced cavitation; Optical breakdown; Thermoelastic acoustic response; Initial bubble geometry; Plasma-thermal coupling

## 1. Introduction

Laser-liquid interactions under pulsed laser irradiation involve a sequence of ultrafast, strongly coupled processes, including rapid optical energy deposition, localized heating, pressure transients, phase change, and fluid motion [1], [2], [3], [4]. As illustrated in **Figure 1(a)**, under nanosecond optical breakdown conditions, nonlinear ionization and plasma formation can be followed by thermal energy deposition and a thermoelastic response within a very short time [1], [3]. These processes lead to the inception of an initial cavity, which subsequently evolves into a bubble. The expansion and collapse of this bubble can generate intense fluid motion, pressure waves, and high-speed liquid jets [5], [6], [7]. Such dynamics are central to a wide range of applications, including laser microsurgery, microfluidic actuation, material transport in MEMS systems, and bioprinting [8], [9], [10]. One representative example is the laser-induced forward transfer (LIFT), as shown in **Fig. 1(b)**, in which bubble expansion drives a jet that propels material from an ink-coated donor layer toward a receiving substrate, thereby enabling material deposition [11], [12]. In this process, the size, shape, and early-time dynamics of the bubble directly influence jet initiation, transferred volume, and deposition stability [11], [13]. Therefore, understanding the physical mechanisms governing bubble inception and early expansion is essential for improving the control and reliability of laser-driven liquid processing.

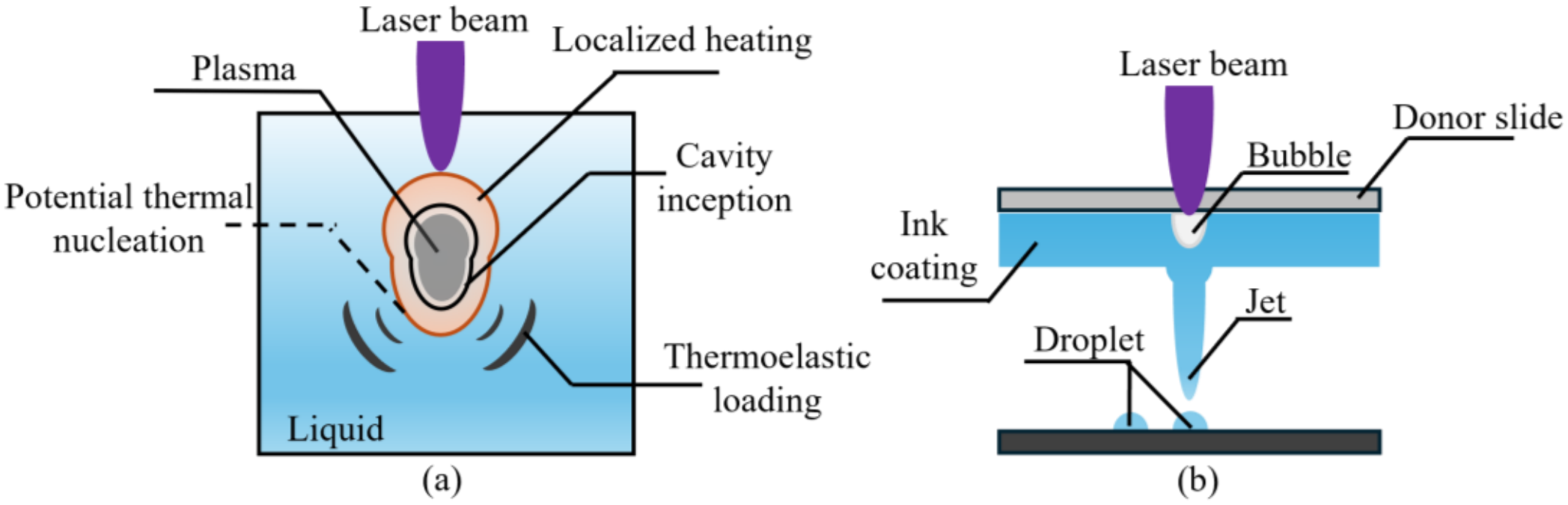

***Fig. 1. Schematic of laser-induced cavitation and its role in LIFT printing.*** *(a) In bulk liquid, focused pulsed irradiation can produce plasma formation, localized heating, thermoelastic loading, and cavity inception, a possible thermal nucleation pathway indicated for comparison. (b) In LIFT printing, bubble expansion within the ink coating drives jet formation and droplet transfer toward the receiving substrate.*

Existing modeling approaches can be broadly categorized into two physically distinct limiting descriptions. First, plasma-related models describe multiphoton ionization, avalanche growth, and inverse Bremsstrahlung absorption in detail, thereby capturing a highly localized and strongly nonlinear energy-deposition pathway in which the surrounding liquid cannot fully relax during energy delivery [1], [2], [3], [14], [15], [16]. In most such models, however, moving breakdown is represented through an effective evolution of plasma length and average absorption, while the spatial distribution of absorption within the plasma is treated in a simplified manner [17], [18], [19]. Consequently, these models do not fully resolve the upstream growth of the breakdown region, the asymmetric plasma shape caused by downstream shielding, or the relationship between deposited energy and the resulting mechanical and thermal responses. Second, thermal nucleation models treat bubble formation as a phase-transition process driven by volumetric superheating, associating the earliest cavity with the local thermodynamic state of the liquid rather than with a resolved plasma-related mechanical response [20], [21], [22]. Related continuum and computational fluid dynamics (CFD) studies extend this thermal nucleation framework to post-formation bubble evolution with increasing thermodynamic detail, including phase change and, in some cases, bubble chemistry [23], [24], [25], [26]. Nevertheless, they generally begin from a prescribed initial cavity state rather than predicting its emergence from optical breakdown [23], [25]. These two modeling descriptions should therefore be regarded as physically meaningful limiting regimes rather than mutually exclusive explanations.

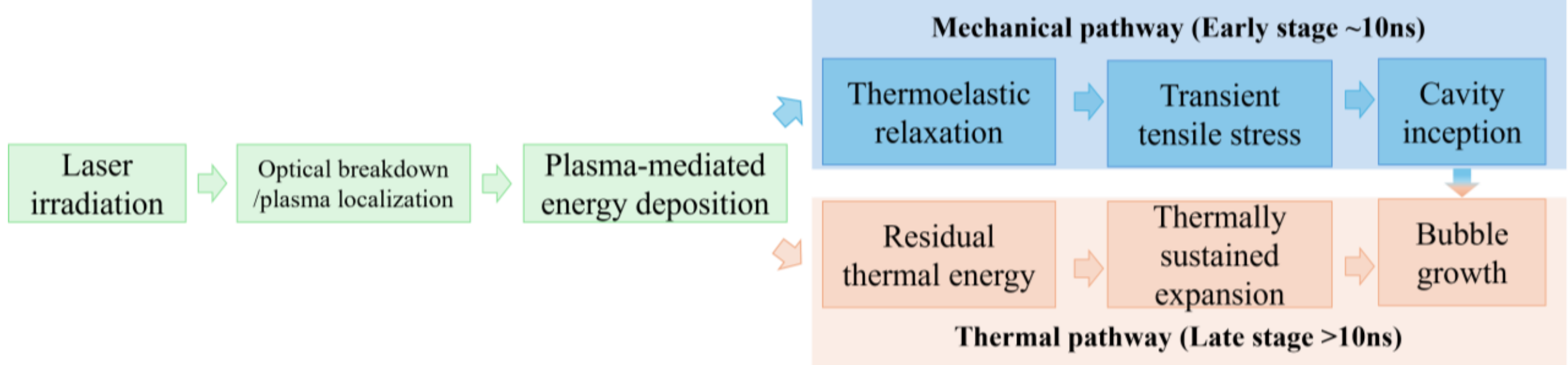

***Figure 2. Conceptual structure of the coupled plasma-thermal framework for laser-induced bubble inception and early growth.*** *Plasma-mediated energy deposition provides a common source for two coupled responses: a mechanical pathway in which thermoelastic acoustic relaxation generates transient tensile loading and initiates bubble inception, and a thermal pathway in which residual heat supports thermally sustained bubble growth.*

This study establishes a unified mechanistic framework for laser-induced bubble inception by resolving the mechanical and thermal pathways as coupled outcomes of a single plasma-mediated energy deposition event. Rather than treating optical breakdown and thermal nucleation as separate limiting descriptions, the present approach explicitly links plasma absorption, transient stress generation, cavity inception, and early bubble growth within one physical picture. In this framework, optical breakdown first generates a localized and anisotropic plasma absorption region, which acts as a common source for two dynamically coupled responses, as summarized in **Fig. 2**. Rapid energy deposition drives a thermoelastic mechanical pathway, producing transient tensile loading that initiates the earliest cavity, while the residual deposited energy forms a thermal pathway that supports subsequent bubble growth. The relative importance of these pathways therefore emerges naturally as the bubble evolves, rather than being prescribed a priori. Importantly, the initial cavity geometry is inherited from the anisotropic plasma distribution, rather than being imposed as a spherically symmetric thermal nucleus. This formulation provides a direct mechanistic bridge between plasma-mediated optical breakdown and the early hydrodynamic state that controls laser-induced cavitation and LIFT jet formation.

Here, we develop a coupled plasma-thermal framework that links optical breakdown, plasma energy deposition, mechanically driven cavity inception, residual heat deposition, and post-inception bubble growth under nanosecond irradiation. The model tests the hypothesis that, *above the optical breakdown threshold, the earliest cavity is initiated by a mechanical pathway in which rapid plasma energy deposition drives thermoelastic acoustic relaxation and produces transient tensile loading, while subsequent bubble growth is sustained by a thermal pathway associated with residual energy deposited by the plasma*. By allowing the inception geometry to emerge from the anisotropic plasma shape rather than prescribing a spherical nucleus, the framework provides a predictive link between breakdown-scale energy deposition and hydrodynamic bubble evolution. The results show that nanosecond laser-induced cavitation is better described as a plasma-seeded, thermally sustained process than as either a purely thermal nucleation event or a purely mechanical plasma response.

## 2. Model formulation

### 2.1 Conceptual framework and limiting descriptions

The present framework describes laser-induced bubble formation as an integrated plasma-thermal process rather than as a set of independent mechanisms. As introduced in **Fig. 2**, plasma-mediated energy deposition serves as the common source for two coupled responses. The first is a mechanical pathway, in which rapid localized energy deposition produces thermoelastic loading that can relax into tensile stress and initiate the earliest cavity. The second is a thermal pathway, in which residual energy retained in the liquid contributes to heat accumulation, vapor generation, and sustained bubble growth after the initial mechanical impulse decays.

Within this framework, the plasma-mediated and thermal-dominated descriptions are treated as limiting cases used to isolate the roles of the two pathways. In the plasma-mediated limit, laser energy is deposited through free-electron generation, cascade ionization, and inverse Bremsstrahlung absorption [1], [3], [16], [17], [19]. The resulting energy deposition is highly localized within the breakdown region and produces a transient thermoelastic pressure response that provides the mechanical route to cavity inception. In the thermal-dominated limit, bubble formation is instead described as a phase-transition process driven by volumetric heating and superheating, with the earliest cavity associated with local thermodynamic nucleation rather than a resolved plasma-driven mechanical response [20], [22].

These two descriptions represent physically meaningful limiting regimes. However, under nanosecond optical breakdown conditions, the mechanical and thermal pathways are not independent. Once plasma absorption occurs, part of the deposited energy generates a transient thermoelastic stress field, while the remaining thermal energy is retained in the liquid and contributes to later bubble growth. The coupled plasma-thermal framework therefore resolves both pathways on a unified spatiotemporal grid, allowing the inception mechanism and post-inception growth to emerge from the same plasma-mediated energy source.

This formulation leads to two important consequences. First, the earliest cavity is not treated as a purely thermal nucleus formed at the superheat limit, but as a mechanically initiated cavity driven by transient tensile stress generated after rapid plasma heating. Second, the initial bubble geometry is not prescribed as spherical. Because the absorbed energy distribution is anisotropic under moving-breakdown and shielding conditions, the nascent cavity inherits the spatial morphology of the plasma. For quantitative comparison with imaging-based measurements, the bubble size is reported later as an equivalent spherical radius, while the underlying model retains the asymmetric energy-deposition physics.

Building on the conceptual structure defined above, the coupled framework is implemented as a staged physical model that links plasma generation, thermoelastic stress generation, cavity inception, and post-inception bubble growth. **Fig. 3** summarizes this physical sequence. Once the focused laser pulse exceeds the optical breakdown condition, free electrons are generated and amplified through multiphoton and cascade ionization, leading to plasma formation and rapid energy deposition within the focal volume. Because this energy is deposited on a timescale shorter than the characteristic hydrodynamic expansion time of the surrounding liquid, the heated region cannot expand instantaneously. The resulting confined thermal expansion gives rise to a transient thermoelastic pressure response, which subsequently evolves into tensile loading capable of initiating an early cavity. After the plasma-driven mechanical impulse decays, the residual thermal energy contributes to post-plasma bubble growth through heat retention, vapor generation, and liquid inertia.

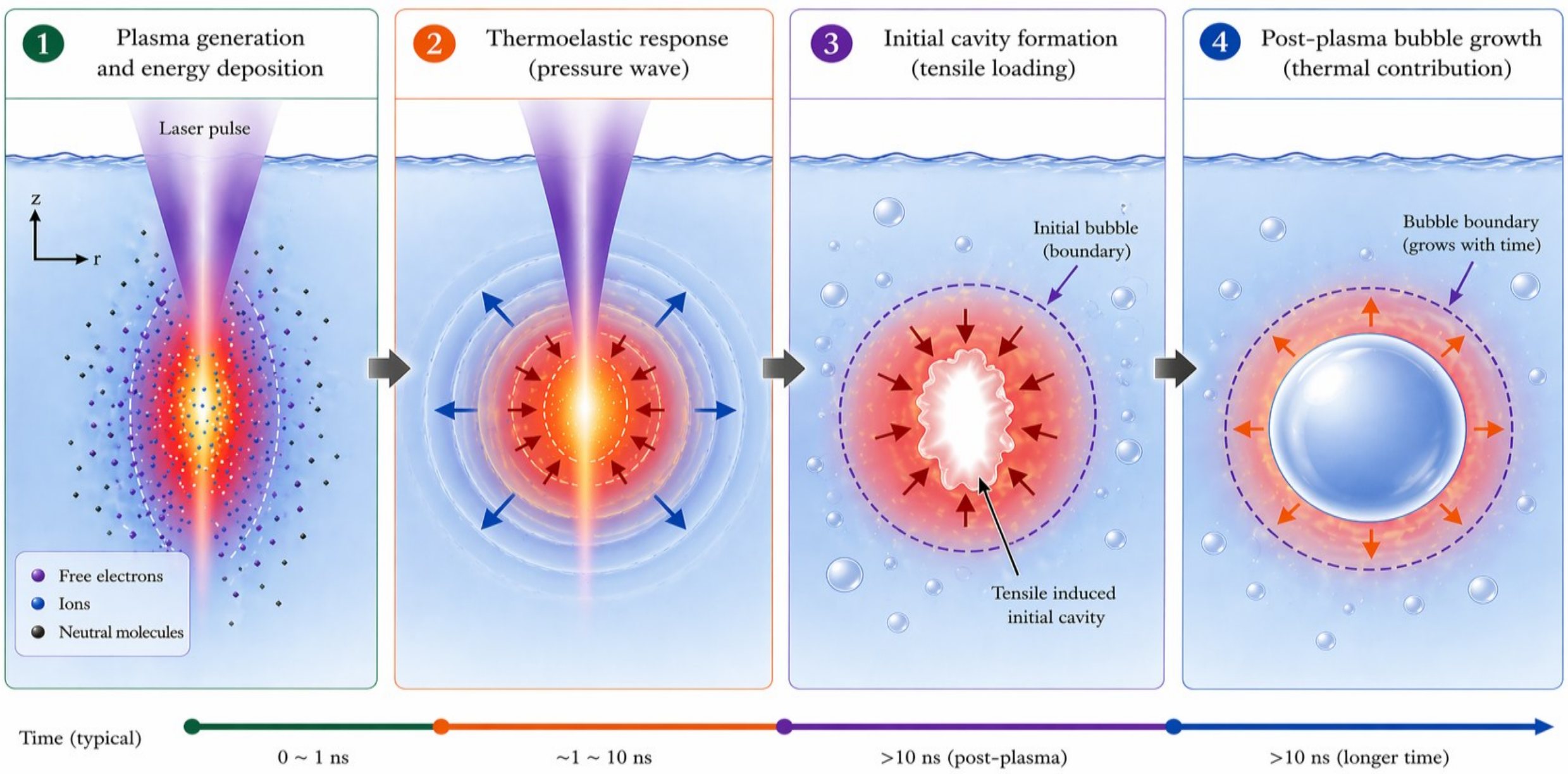

***Figure 3. Staged physical sequence represented in the coupled plasma-thermal framework.*** *Plasma generation and localized energy deposition first produce a thermoelastic source pressure, whose acoustic relaxation generates tensile loading and cavity inception; residual thermal energy then contributes to post-plasma bubble growth.*

The following subsections formulate the two limiting cases and the coupled plasma-thermal model, beginning with the plasma-mediated limit that governs breakdown-scale energy deposition and the initial thermoelastic response.

## 2.2 Plasma-mediated limiting case

The plasma-mediated limiting case isolates the breakdown and plasma driven pathway in the coupled framework. In this limit, laser energy deposition is governed by free electron generation, plasma absorption, and shielding induced localization, while thermal nucleation is not used as the inception criterion. The purpose of this limiting case is to quantify how much of the observed bubble inception can be explained by plasma-mediated energy deposition and the resulting thermoelastic loading alone.

The laser beam is modeled as a Gaussian pulse in both space and time[1], [27]. The intensity distribution is expressed as

$$I(r,z,t) = I_0 \left(\frac{w_0}{w(z)}\right)^2 \exp\left(-\frac{2r^2}{w^2(z)}\right) \exp\left(-\frac{4\,ln2 \cdot t^2}{\tau_L^2}\right) \tag{1}$$

where $w_0$ is the beam waist radius, $\tau_L$ is the pulse duration (FWHM), and $w(z)$ is the beam radius along the propagation direction given by:

$$w(z) = w_0\sqrt{1 + \left(\frac{z}{z_R}\right)^2} \tag{2}$$

Here, $z_R$ is the Rayleigh length:

$$z_R = \frac{\pi n_0 w_0^2}{\lambda} \tag{3}$$

$n_0$ is the refractive index of water, and λ is the laser wavelength.

The evolution of the free electron density $\rho$ is described by a rate equation accounting for electron generation and loss mechanisms[1], [2], [3], [17],

$$\frac{d\rho}{dt} = W_{MPI} + W_{avalanche}\,\rho - W_{diff}\,\rho - \eta_{rec}\,\rho^2 \tag{4}$$

The first term represents multiphoton ionization (MPI), which initiates the breakdown process. The second term corresponds to cascade (avalanche) ionization driven by inverse Bremsstrahlung absorption. The last two terms account for electron diffusion out of the focal volume and bimolecular recombination, respectively.

This formulation follows the classical description of laser-induced breakdown in liquids and captures the transition from MPI-dominated initiation to avalanche-dominated growth during nanosecond pulses.

Once free electrons are generated, laser energy is absorbed primarily through inverse Bremsstrahlung [14], [28]. The corresponding plasma absorption coefficient is written as

$$\alpha_{IB} = \frac{\rho e^2 \nu_c}{c n_0 m_e \varepsilon_0 (\omega^2 + \nu_c^2)} \tag{5}$$

where $\nu_c$ is the effective collision frequency, $\omega$ is the laser angular frequency.

The attenuation of laser intensity along the propagation direction is governed by

$$\frac{dI}{dz} = -\alpha_{IB}\, I \tag{6}$$

which naturally accounts for plasma shielding and the moving breakdown effect.

The absorbed energy density is obtained as

$$u_{Heat}(r,z) = \int \alpha_{IB} I(r,z,t)dt \tag{7}$$

This quantity represents the local energy deposited into the liquid by the plasma.

The deposited energy produces a local temperature rise

$$\Delta T = \frac{u_{heat}}{\rho_L c_p} \tag{8}$$

where $\rho_L$ is the liquid density and $c_p$ is the specific heat capacity.

The rapid temperature rise first defines a local thermoelastic source pressure,

$$p_0(r,z) = \frac{\beta}{\kappa_T}\Delta T(r,z) \tag{9}$$

where the thermal expansion coefficient $\beta$ and isothermal compressibility of water $\kappa_T$ determine the source-pressure scale. Equivalently, this can be written as the product of bulk modulus, thermal expansion coefficient, and temperature rise $p_0 = K\beta\Delta T$, where $K = 1/\kappa_T$ is the bulk modulus. This quantity represents the initially compressive source-pressure scale generated by localized plasma heating. It is not used directly as the tensile cavitation pressure; the tensile phase arises from acoustic relaxation of this compressive source, as described below.

The dynamic thermoelastic pressure is then obtained by propagating this source through an axisymmetric acoustic response model [29].

$$\frac{\partial^2 p}{\partial t^2} - c_s^2 \left[\frac{\partial^2 p}{\partial r^2} + \frac{1}{r}\frac{\partial p}{\partial r} + \frac{\partial^2 p}{\partial z^2}\right] = S(r, z, t) \tag{10}$$

where $p(r,z,t)$ is the time-dependent pressure field, and $c_s$ is the speed of sound in water.

For nanosecond heating, the source term is represented as

$$S(r, z, t) = p_0(r, z)\frac{dg(t)}{dt} \tag{11}$$

where $g(t)$ is a normalized Gaussian temporal heating function with full width at half maximum equal to the laser pulse duration.

This source term $S(r, z, t)$ separates the spatial energy-deposition profile from the temporal heating envelope. The approximation is used because the thermoelastic response is governed by the integrated plasma absorption field, while the temporal energy delivery occurs over approximately one pulse duration. The characteristic propagation speed of the moving breakdown front is on the order of $10^6$ to $10^7$ m/s, while the acoustic propagation speed is $c_s \approx 1500$ m/s [30]. The plasma spatial redistribution therefore occurs on a timescale much shorter than the acoustic relaxation time, so that by the time acoustic propagation becomes significant, the spatial energy deposition profile has already reached its final configuration. This treatment accounts for finite-duration thermoelastic loading rather than assuming an instantaneous fully stress-confined pressure jump.

As the initially compressive pressure pulse propagates away from the focal region, a local rarefaction phase develops and produces the transient tensile pressure used for mechanical cavitation. Mechanical cavitation is assumed to occur when the dynamic thermoelastic pressure satisfies

$$p(r, z, t) \leq -p_{cav} \tag{12}$$

where $p_{cav} > 0$ denotes the magnitude of the tensile cavitation threshold. The threshold $p_{cav} = 100$ MPa is representative of the dynamic tensile strength of water under nanosecond stress confinement, consistent with values reported in photoacoustic cavitation studies [19] and laser breakdown experiments [20].

### 2.2.1 Bubble inception and shape initialization

After the dynamic thermoelastic pressure field is obtained, the mechanically activated inception region is identified from the strongest tensile rarefaction reached during acoustic relaxation. The local minimum

pressure is defined as

$$p_{min}(r,z) = \min_t p(r,z,t) \tag{13}$$

The initial cavity supporting region is then defined as the set of points where $p_{min}(r,z)$ is less than or equal to $-p_{cav}$.

$$\Omega_{b,0} = \{(r,z) \in \Omega : p_{min}(r,z) \leq -p_{cav}\} \tag{14}$$

This definition links the inception region directly to the dynamic tensile pressure field rather than to the initial compressive source pressure. For the axisymmetric configuration considered in the present study, the initial bubble volume is evaluated as

$$V_{\mathrm{b},0} = 2\pi \iint_{\Omega_{\mathrm{b},0}} r\, dr\, dz \tag{15}$$

The corresponding effective initial bubble size is then expressed through an equivalent spherical radius,

$$R_0 = \left(\frac{3V_{\mathrm{b},0}}{4\pi}\right)^{\frac{1}{3}} \tag{16}$$

This definition provides a consistent scalar measure for comparison with experimentally reported bubble radii, while not implying that the onset cavity is spherical. In the present work, $R_0$ also serves as the initialization radius for the post-inception $R(t)$ calculation described in **Sec. 2.5.** The underlying anisotropic inception physics is characterized separately through the geometric descriptors introduced below.

Because the initial cavity in the present model is inherited from the plasma morphology, its shape is further characterized by an elongation factor and an axial asymmetry factor. Let $z_c$ denote the characteristic axial center of the initial cavity, and let $L_f$ and $L_b$ denote the forward and backward axial extents relative to $z_c$, while $L_r$ denotes the maximum radial extent. The elongation factor is defined as

$$\chi = \frac{L_{\mathrm{f}} + L_{\mathrm{b}}}{2L_{\mathrm{r}}} \tag{17}$$

and the axial asymmetry factor is defined as

$$\delta = \frac{L_{\mathrm{f}} - L_{\mathrm{b}}}{L_{\mathrm{f}} + L_{\mathrm{b}}} \tag{18}$$

Here, $\chi$ quantifies the overall axial stretching of the onset cavity, while $\delta$ distinguishes a front-back symmetric ellipsoidal structure from a pear-like cavity with directional skewness. In this way, the initial bubble is described by three complementary quantities: the effective size $R_0$, the global elongation $\chi$, and the axial asymmetry $\delta$. This representation preserves the geometric signature of asymmetric plasma-mediated energy deposition while remaining sufficiently compact for subsequent bubble-dynamics analysis.

### 2.3 Thermal-only limiting case

The thermal-only limiting case isolates the phase-transition pathway in the absence of explicitly resolved plasma kinetics. In this limit, laser energy deposition is treated as effective volumetric heating, and bubble inception is determined by thermally activated nucleation rather than by a plasma-driven tensile stress criterion. This case is used diagnostically to evaluate how a purely thermal description performs when applied to nanosecond breakdown conditions.

The absorbed energy density is expressed as

$$u_{heat}^{th}(r,z) = \int \alpha_{eff}\, I(r,z,t)\, dt \tag{19}$$

where $\alpha_{eff}$ is an effective absorption coefficient that represents the net conversion of laser energy into thermal energy. Here, $\alpha_{eff}$ is not interpreted as a microscopic absorption coefficient. Instead, it represents the effective volumetric heating strength required to match the total absorbed energy of the plasma-mediated case. To avoid biasing the thermal-only limiting case through an arbitrary absorption strength, $\alpha_{eff}$ is determined by absorbed energy matching. For each laser energy, the total thermal energy deposited in the thermal-only model is matched to the total plasma-mediated absorbed energy. Thus, the thermal-only model is evaluated under the same total absorbed-energy constraint as the plasma-mediated model.

$$\int_V u_{heat}^{th}\, dV = \int_V u_{heat}^{pl}\, dV \tag{20}$$

Differences between the two limiting cases, therefore, arise from the spatial distribution of deposited energy and the inception mechanism, rather than from prescribing different total absorbed energies. In contrast to the plasma-mediated case, the energy distribution is less localized, resulting in a smoother temperature and pressure field.

The resulting temperature rise follows the same relation as in the plasma-mediated model,

$$\Delta T = \frac{u^{th}\, heat}{\rho_L c_p} \tag{21}$$

From a thermal perspective, rapid heating may drive the liquid into a metastable superheated state, in which vapor nuclei forms through classical nucleation theory (CNT) [21], [22]. The nucleation rate is written as

$$J(T) = J_0 \exp\left(-\frac{\Delta G^*}{k_B\, T}\right) \tag{22}$$

where $J_0$ is the kinetic prefactor, $k_B$ is the Boltzmann constant, and $\Delta G^*$ is the free-energy barrier for the formation of a critical nucleus.

The free energy barrier is expressed as [21], [22]

$$\Delta G^* = \frac{16\pi\sigma^3}{3(P_v - P_L)^2} \tag{23}$$

where $\sigma$ is the surface tension, $P_v$ is the vapor pressure inside the embryo, and $P_L$ is the surrounding liquid pressure.

To account for the temperature dependence of surface tension, $\sigma(T)$ is evaluated using[31]

$$\sigma(T) = 0.2358\, \tau^{1.256}(1 - 0.625\tau)\,, \quad \tau = 1 - \frac{T}{T_c} \tag{24}$$

where $T_c$ is the critical temperature of water.

The vapor pressure is taken as a temperature-dependent quantity. In the present model, nucleation is assumed to become significant when the cumulative nucleation probability exceeds unity within the heated volume,

$$N_{cum} = \int_{t_0}^{t} \int_V J(T(x,t))\, dt\, dV \geq 1 \tag{25}$$

In the present framework, this CNT-based criterion represents the thermal limiting pathway and also provides a mechanism by which residual heat can contribute to vapor generation during later bubble growth. For the thermal-only model, bubble inception is assumed to occur when the cumulative nucleation criterion is first satisfied, i.e.,

$$t_{\mathrm{inc}}^{(\mathrm{th})} = inf\,\{t | N_{\mathrm{cum}}(t) \geq 1\} \tag{26}$$

At $t = t_{\mathrm{inc}}^{(\mathrm{th})}$, the initial thermally activated bubble-supporting region is defined as

$$\Omega_{\mathrm{b},0}^{(\mathrm{th})} = \left\{ \mathbf{x} \in \Omega \middle| T\left(\mathbf{x},\, t_{\mathrm{inc}}^{(\mathrm{th})}\right) \geq T_{\mathrm{nuc}} \right\} \tag{27}$$

where $T_{\mathrm{nuc}}$ denotes the local temperature at which the nucleation criterion becomes significant under the present CNT-based formulation.

For the axisymmetric configuration, the corresponding initial bubble volume is evaluated as

$$V_{\mathrm{b},0}^{(\mathrm{th})} = 2\pi \iint_{\Omega_{\mathrm{b},0}^{(\mathrm{th})}} r\, dr\, dz \tag{28}$$

The effective initial bubble radius is then defined as

$$R_0^{(\mathrm{th})} = \left( \frac{3V_{\mathrm{b},0}^{(\mathrm{th})}}{4\pi} \right)^{\frac{1}{3}} \tag{29}$$

This equivalent spherical radius provides a consistent scalar measure for comparison with experimentally reported bubble radii, while not implying that the thermally activated cavity is strictly spherical. In the present work, $R_0^{(\mathrm{th})}$ is likewise used as the initialization radius for the post-inception $R(t)$ calculation described in **Sec. 2.5**.

### 2.4 Coupled plasma-thermal formulation

The coupled plasma-thermal formulation represents the proposed model. Unlike the two limiting cases, it does not assign bubble formation to a single mechanism. Instead, plasma-mediated energy deposition is treated as a common source that drives two coupled responses: a thermoelastic mechanical response that controls the earliest cavity inception, and a residual thermal response that contributes to post-inception bubble growth.

Here, the term coupled denotes that the mechanical and thermal responses are driven by the same plasma-mediated energy deposition field and are evaluated within a common spatiotemporal framework.

The present model does not resolve full two-way feedback among bubble growth, pressure-field evolution, and plasma kinetics after inception. Instead, it provides a staged common source coupling that links breakdown-scale energy deposition to both mechanical inception and thermally sustained growth.

The free electron density is first obtained from the rate equation in **Sec. 2.2**, and the corresponding absorbed energy field is calculated through inverse Bremsstrahlung absorption. Rapid deposition within the breakdown region generates a localized temperature rise and an associated thermoelastic pressure field. The compressive pressure response subsequently relaxes into a tensile field, providing the mechanical criterion for early cavity inception.

At the same time, residual heat remains in the surrounding liquid after the plasma driven mechanical impulse decays. This retained thermal energy contributes to superheating, vapor generation, and sustained bubble growth, following the CNT-based thermal description introduced in **Sec. 2.3**.

In the coupled formulation, the inception time is defined as the earliest time at which either the mechanical tensile criterion or the thermal nucleation criterion is satisfied:

$$p < -p_{cav} \tag{30}$$

or

$$N_{cum} = \int_V \int_t J(T)\, dt\, dV \geq 1 \tag{31}$$

The corresponding birth state is then constructed from the active inception region and mapped to an equivalent spherical radius $R_0^{(c)}$, which is used to initialize the post-inception radius evolution in **Sec. 2.5**. This staged common source coupling captures the transition from pressure-driven inception to thermally sustained growth, providing a continuous description of laser-induced bubble formation.

**2.5 Post-inception bubble dynamics**

After inception, each formulation provides an initial bubble state, while the subsequent radius evolution is evaluated using a common continuum-scale model. This separation ensures that differences among the plasma-mediated, thermal-only, and coupled predictions arise from the birth state rather than from different post inception dynamics.

For each formulation, $R_0$ is obtained from the corresponding inception volume defined in the preceding sections. The initial vapor pressure contribution $P_{v,0}$ is evaluated from the average temperature within the inception region using the same temperature dependent vapor pressure relation used in the thermal nucleation calculation. The initial internal bubble pressure $P_{B,0}$ is initialized from the local thermodynamic state of the inception region together with the capillary pressure associated with $R_0$. Because the three formulations predict different inception temperatures and cavities supporting regions, the resulting $P_{v,0}$ and $P_{B,0}$ can differ among models. These differences are treated as physical consequences of the predicted birth state rather than as independently prescribed Rayleigh-Plesset parameters.

The use of an equivalent spherical Rayleigh-Plesset model does not imply that the inception cavity is assumed to be spherical. The anisotropic inception region is first used to determine the birth volume and equivalent radius, while the subsequent radius evolution is represented by a reduced-order scalar model for comparison with imaging-derived equivalent radius. The detailed non-spherical interface dynamics after inception are not resolved in the present formulation and require axisymmetric or three-dimensional CFD treatment.

The subsequent bubble expansion is represented through the evolution of an equivalent spherical bubble radius $R(t)$. In the present study, this post-inception stage is evaluated using the Rayleigh-Plesset equation for an equivalent spherical bubble [32], [33], [34],

$$\rho_{\mathrm{L}}\left(R\ddot{R} + \frac{3}{2}\dot{R}^2\right) = P_{\mathrm{B}} - P_\infty - \frac{2\sigma}{R} - \frac{4\mu\dot{R}}{R} \tag{32}$$

where $\rho_L$ is the liquid density, $P_B$ is the effective internal bubble pressure, $P_\infty$ is the ambient liquid pressure, $\sigma$ is the surface tension, and $\mu$ is the liquid viscosity.

The internal bubble pressure is approximated by a polytropic relation [34]

$$P_{\mathrm{B}}(R) = P_{\mathrm{v},0} + \left(P_{\mathrm{B},0} - P_{\mathrm{v},0}\right)\left(\frac{R_0}{R}\right)^{3\kappa_0} \tag{33}$$

where $R_0$ is the equivalent inception radius, $P_{B,0}$ is the initialized internal bubble pressure, $P_{v,0}$ is the vapor-pressure floor, and $\kappa_b$ is the effective polytropic exponent.

For the plasma-mediated, thermal-only, and coupled formulations, the later $R(t)$ evolution is evaluated using the same Rayleigh-Plesset framework, while the differences among the three model predictions arise from how the inception state ($R_0$, $P_{B,0}$, $P_{v,0}$) is established.

**Table 1.** Model parameters used in this study.

| Parameter | Symbol | Value | Source |
|---|---|---|---|
| ***Laser parameters*** | | | |
| Laser wavelength | $\lambda$ | 1064 nm | [4] |
| Pulse duration (FWHM) | $\tau_{\mathrm{L}}$ | 10 ns | [4] |
| Beam waist | $w_0$ | 3.25 μm | [4] |
| ***Liquid properties (water at 293 K, 0.1 MPa)*** | | | |
| Density | $\rho_{\mathrm{L}}$ | 1000 kg/m³ | standard |
| Specific heat capacity | $c_{\mathrm{p}}$ | 4180 J/(kg·K) | standard |
| Surface tension | $\sigma$ | 0.072 N/m | [31] |
| Dynamic viscosity | $\mu$ | 0.001 Pa·s | standard |

| Thermal expansion coeff. | $\beta$ | $2.57 \times 10^{-4}$ K$^{-1}$ | [21], [22] |
|---|---|---|---|
| Isothermal compressibility | $\kappa_T$ | $4.6 \times 10^{-10}$ Pa$^{-1}$ | [21], [22] |
| Sound speed | $c_s$ | 1500 m/s | standard |
| Thermal diffusivity | $\alpha_{th}$ | $1.4 \times 10^{-7}$ m$^2$/s | standard |
| ***Thermoelastic acoustic response*** | | | |
| Source coefficient | $\Gamma = K\beta/(\rho_L c_p)$ | ~0.13 | derived [a] |
| Tensile cavitation threshold | $p_{cav}$ | $1.0 \times 10^8$ Pa | [1], [3] [b] |
| ***Plasma kinetics*** | | | |
| Recombination coefficient | $\eta_{rec}$ | $2.0 \times 10^{-9}$ cm$^3$/s | [1] |
| MPI cross section (8-photon) | $\sigma_8$ | see [3] | [3] |
| Cascade ionization rate | $\eta_{casc}$ | see [3] | [3] |
| ***Thermal nucleation (CNT)*** | | | |
| Nucleation prefactor | $J_0$ | $1.55 \times 10^{38}$ m$^{-3}$s$^{-1}$ | [21], [22] |
| Surface tension relation | $\sigma(T)$ | Eq. (24) | [31] |
| ***Thermal-only limiting case*** | | | |
| Eff. absorption coefficient | $\alpha_{eff}$ | energy-matched | Eq. (20) [c] |

[a] The thermoelastic source coefficient Γ is not an independent parameter. It is evaluated as $\Gamma = K\beta/(\rho_L c_p) = \beta/(\kappa_T \rho_L c_p)$ using the tabulated water properties at ambient conditions. The thermoelastic parameters $\beta$ and $\kappa_T$ are taken from the photoacoustic and laser-ablation literature [2], [35]

[b] The threshold $p_{cav}$ = 100 MPa represents the dynamic tensile strength of water under nanosecond stress confinement, consistent with photoacoustic cavitation studies [1] and laser breakdown experiments [3]. The effective tensile strength depends on water purity, dissolved gas content, confinement duration, and nucleation-site availability. The same value is applied across all energies and model formulations.

[c] The effective absorption coefficient $\alpha_{eff}$ is determined by absorbed-energy matching (Eq. 20), ensuring that the thermal-only model receives the same total absorbed energy as the plasma-mediated model at each laser energy. Differences between limiting cases arise from the spatial distribution of deposited energy, not from different total energy inputs.

## 3. Results and Discussion

### 3.1 Early time free electron dynamics and ionization pathways

The temporal evolution of the free electron density provides a baseline test for the plasma kinetic component of the model. Figure 4 compares the simulated electron density histories for three representative pulse durations with the benchmark calculations of Noack and Vogel [1]. Time is normalized by the pulse duration $\tau_L$, with t = 0 defined at the pulse peak; negative and positive values of $t/\tau_L$ therefore denote times before and after the pulse peak, respectively.

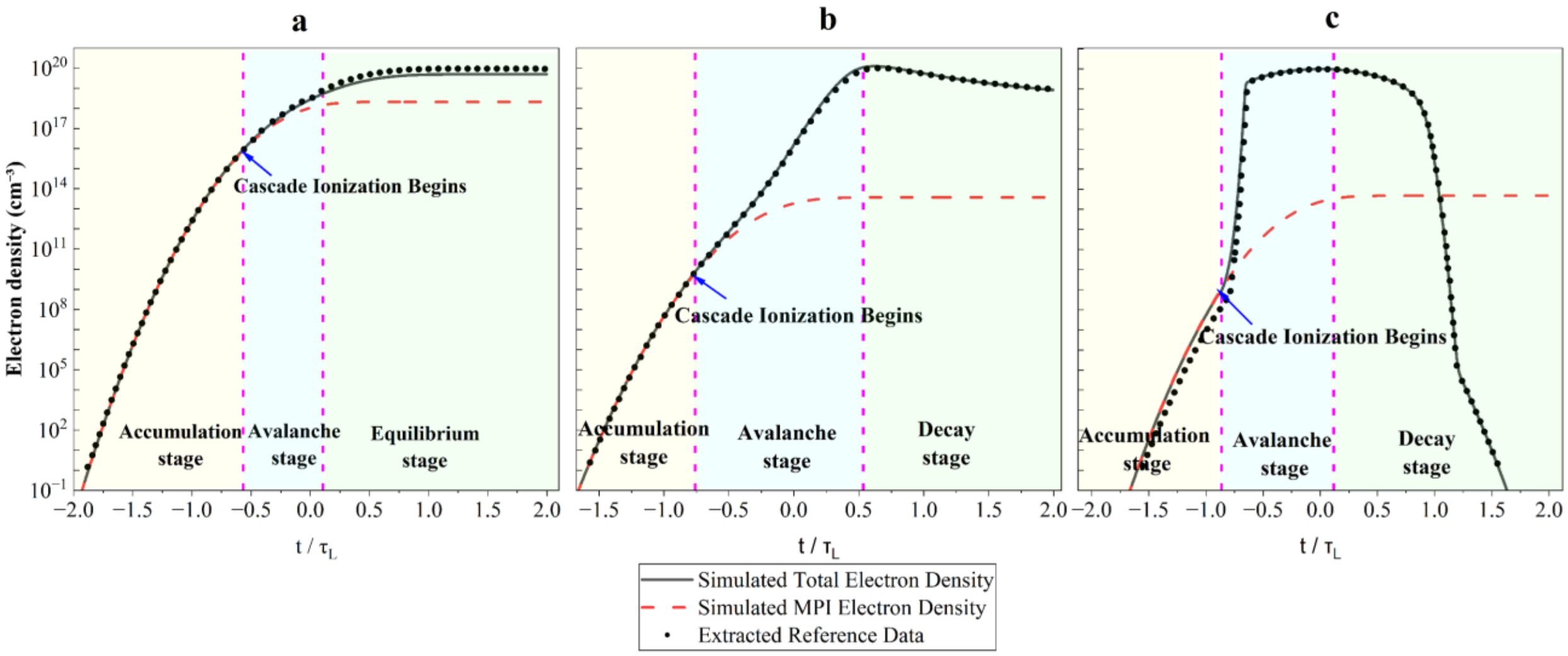

***Fig. 4. Temporal evolution of free electron density for representative laser pulse durations.*** *Simulated total electron density and MPI only contribution is compared with benchmark calculations for* ***(a)*** $\tau_L = 100fs$, $I = 7.7 \times 10^{12} W/cm^2$, ***(b)*** $\tau_L = 30ps$, $I = 1.2 \times 10^{11} W/cm^2$, *and* ***(c)*** $\tau_L = 6ns$, $I = 3.4 \times 10^{10} W/cm^2$. *Time is normalized by the pulse duration, with* $t/\tau_L = 0$ *corresponding to the pulse peak. The deviation between the total and MPI only curves identifies the onset of avalanche-dominated growth.*

From **Fig. 4**, the proposed model accurately captured the three early stages of optical breakdown in water: (1) initial seed-electron generation by multiphoton ionization (MPI); (2) the onset of cascade (avalanche) ionization, and (3) subsequent saturation or decay governed by recombination and diffusion losses. At early times, before the pulse peak, the electron density increases gradually as MPI provides the initial seed population. As the laser intensity approaches the pulse peak, inverse Bremsstrahlung absorption by seed electrons drives cascade ionization, producing rapid multiplication of free carriers. For the nanosecond condition relevant to the present study, this avalanche-driven growth occurs over approximately 2-3 ns around the pulse peak. The transition from MPI-dominated seeding to avalanche-dominated growth is identified by the point where the total electron density curve departs from the MPI-only curve.

The pulse-duration dependence of the ionization pathways reflects fundamental differences in the relative roles of MPI and cascade processes. In the femtosecond regime (**Fig. 4a**), the pulse terminates before significant avalanche multiplication can develop. As a result, MPI dominates throughout the pulse, the electron density reaches a well-defined plateau, and only minimal post-pulse decay is observed. This behavior is consistent with the deterministic, intensity-dependent character of femtosecond breakdown described by Vogel et al. [3], [28], where the short interaction time precludes appreciable collisional energy gain by conduction-band electrons. In contrast, picosecond and nanosecond pulses (**Fig. 4b-c**) provide sufficient interaction time for cascade ionization to become the dominant electron production mechanism, consistent with classical descriptions of nanosecond optical breakdown [1], [28]. This is manifested as a sharp exponential rise in electron density near the pulse peak, followed by a pronounced decay phase driven by recombination and diffusion losses after the laser energy input ceases.

The nanosecond regime is of particular significance for the present work. The distinct rise-and-decay profile of the electron density (**Fig. 4c**) indicates that plasma formation is temporally confined to a window of approximately one pulse duration, after which recombination rapidly depletes the free-electron population. This temporal confinement produces a concentrated burst of energy deposition that precedes the longer-timescale thermal and hydrodynamic response of the surrounding liquid. The resulting separation of timescales, sub-nanosecond plasma kinetics versus nanosecond-to-microsecond bubble dynamics, provides the physical basis for the sequential coupling strategy adopted in the present model: plasma-driven energy deposition governs the initial mechanical impulse, while residual thermal energy sustains subsequent bubble evolution.

It should be noted that this temporal validation is performed under the benchmark conditions reported by Noack and Vogel [1], rather than the specific experimental configuration considered in the subsequent analysis. The agreement confirms the fidelity of the rate-equation formulation and the associated ionization parameters, establishing the model as a reliable basis for the spatially resolved analysis that follows.

### 3.2 Plasma morphology and spatially localized absorption

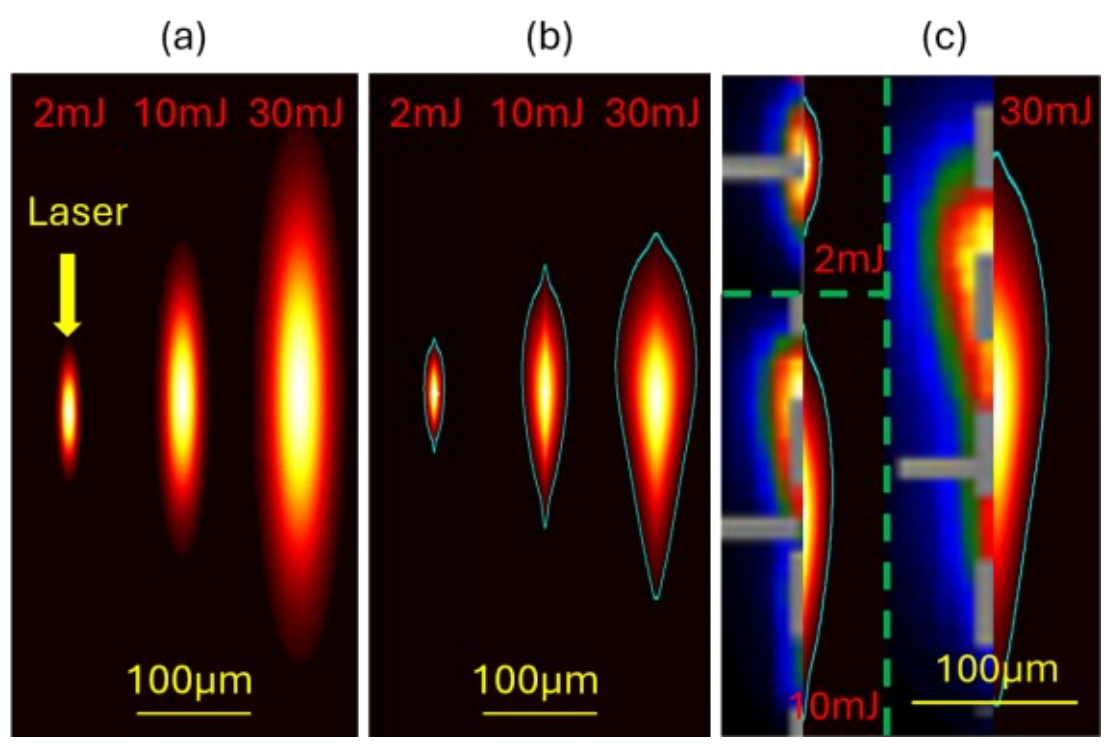

***Fig. 5. Plasma localization and moving-breakdown-induced morphology under nanosecond irradiation.*** *Simulated plasma distribution* ***(a)*** *without plasma shielding and* ***(b)*** *with shielding and moving breakdown included.* ***(c)*** *Comparison with experimental ICCD emission images from Jia et al. [4], showing that shielding-driven upstream energy deposition produces the elongated plasma morphology observed experimentally.*

After the temporal electron-density response is established, the next question is whether the model can reproduce the spatial localization and asymmetry of the plasma. These features determine where laser energy is absorbed and how the initial thermoelastic loading is distributed. Under nanosecond focusing conditions, the plasma morphology is not determined solely by the incident beam geometry; rather, it is shaped by a dynamic feedback mechanism between plasma generation and laser propagation. To assess whether this feedback is correctly captured, the model is applied to the experimental configuration of Jia et al. [4], which provides spatially and temporally resolved measurements of plasma evolution in water at energies ranging from 2 to 30 mJ.

**Fig. 5** presents the simulated spatial distribution of the free-electron density under two conditions: (a) without plasma shielding, and (b) with plasma shielding and the associated moving breakdown effect included. In the absence of shielding (**Fig. 5a**), the electron density distribution remains nearly symmetric about the focal plane, reflecting the Gaussian envelope of the incident beam. The plasma volume is spatially compact and confined to the immediate vicinity of the beam waist. It should be noted that the experimental images from Jia et al. are normalized color ICCD emission maps. The blue/cyan outer structures correspond to a low-intensity plasma emission or a weak emission halo due to colormap normalization, rather than a sharply defined plasma-liquid interface. Therefore, the comparison here is based primarily on the high-emission plasma core and its elongated axial morphology.

When plasma shielding is incorporated (**Fig. 5b**), the predicted plasma emission region became axially elongated and asymmetric. As the electron density builds up within the focal region, inverse Bremsstrahlung absorption increases the local absorption coefficient, attenuating the transmitted laser intensity downstream while simultaneously enhancing energy deposition upstream of the focal plane. This self-reinforcing process shifts the locus of peak energy deposition progressively toward the incoming laser, a well-documented phenomenon known as moving breakdown [1], [3]. The net effect is a pronounced axial elongation of the plasma, with its major axis oriented along the beam propagation direction and an asymmetric concentration of electron density toward the laser source. This conical, upstream elongated morphology is a hallmark of nanosecond breakdown in liquids and has been consistently observed in experimental imaging studies [4], [15].

Across the three laser energies, the predicted plasma lengths are on the same order as the experimentally observed values and show the same increasing trend with input energy. At 10 mJ, the predicted axial plasma extent is approximately 93.4 μm, compared with 88.2 μm extracted from Jia et al. [4]. Similar agreement is obtained at 2 and 30 mJ. Comparison with the fast-imaging measurements of Jia et al. [4] indicates that the inclusion of plasma shielding and moving breakdown is necessary to recover the experimentally observed elongated morphology. This agreement is significant for two reasons. First, it validates the self-consistent treatment of laser propagation and plasma absorption implemented in **Eq. (6)**, where the beam attenuation coefficient evolves dynamically with the local electron density rather than being prescribed a priori. Second, it establishes that the spatial distribution of energy deposition and, consequently, the geometry of the resulting pressure and temperature fields are fundamentally governed by plasma-laser coupling effects that cannot be neglected under nanosecond breakdown conditions.

The elongated plasma pattern marks where laser energy is deposited. This deposition occurs before the surrounding liquid can mechanically respond. Once shielding and moving breakdown extend the absorption region along the beam axis, the resulting thermoelastic pressure field can no longer be treated as radially symmetric. The early mechanical loading therefore inherits the plasma's spatial bias, producing an anisotropic driving field for cavity inception. This interpretation provides a mechanistic explanation for the non-spherical initial bubble shapes observed in nanosecond breakdown experiments [4], [15], and shows why a spatially resolved plasma model is required before a realistic bubble initial condition can be prescribed.

This interpretation also connects the present results to a broader experimental picture of laser-induced cavitation in water. In the references, high-resolution XFEL imaging has shown that plasma-to-cavity transition may proceed through elongated filaments, multiple breakdown sites, asymmetric cavity growth, and collective expansion, rather than through a single spherical energy source [24]. Recent review work similarly emphasizes that plasma formation, shock generation, cavitation, and energy partitioning are coupled stages of one physical sequence rather than separable events [36]. The contribution of the present model is to convert this qualitative picture into a predictive framework for nanosecond breakdown: the model links the spatially resolved plasma absorption field to the birth state of the bubble used in the subsequent continuum-scale evolution.

The following section examines the temperature and pressure fields generated by this asymmetric energy deposition. These fields demonstrate that plasma dynamics do not simply occur before cavitation; they define the spatial origin of the thermoelastic response and determine the geometry from which the earliest bubble emerges.

### 3.3 Thermoelastic fields generated by plasma absorption

Plasma absorption converts the elongated breakdown region into a localized thermal and mechanical source. **Fig. 6** shows the temperature rise and thermoelastic pressure field generated immediately after nanosecond energy deposition for the 10 mJ case. The temperature field is concentrated within the upstream-elongated plasma region identified in **Sec. 3.2**, rather than within a spherical focal volume. The corresponding field represents the thermoelastic source pressure $p_0(r,z)$ calculated from the temperature rise using Eq. (9), and reaches gigapascal-scale peak values near the focal region.

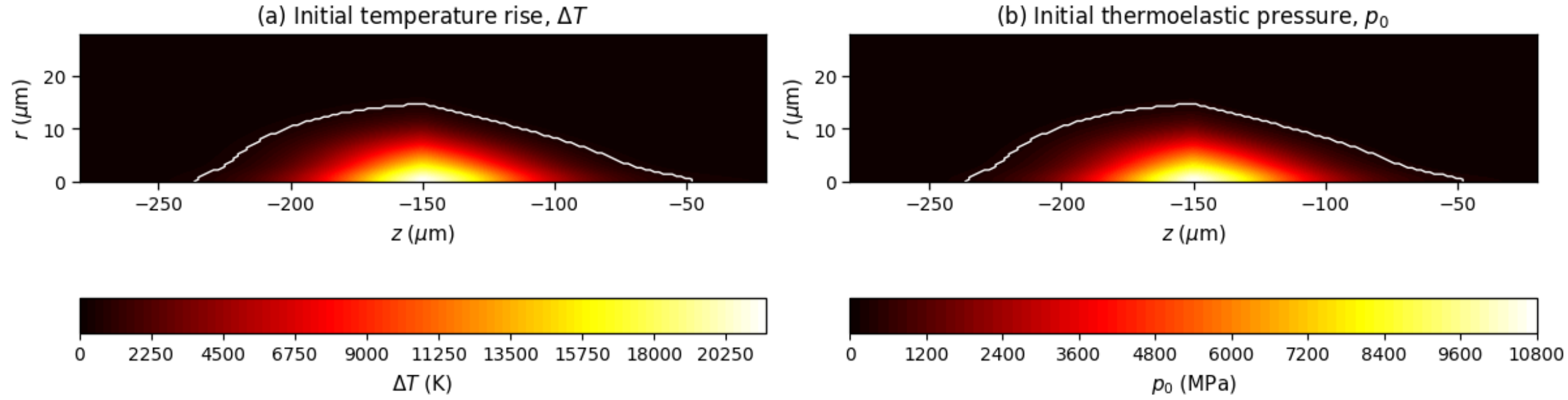

***Fig. 6. Thermoelastic fields generated by plasma-mediated energy deposition.*** *Spatial distributions of (a) temperature rise and (b) compressive thermoelastic source pressure for the 10 mJ case. The white contour marks the plasma-shaped absorption region, showing that both thermal and mechanical fields inherit the elongated morphology of the breakdown region rather than forming around a spherical heat source.*

The pressure field shown in Fig. 6 represents the compressive source-pressure scale $p_0(r,z)$ defined in **Sec. 2.2**, rather than the full time-dependent cavitation pressure. The dynamic pressure evolution resulting from acoustic relaxation of this source field is examined in **Sec. 3.4**. The important feature is not only the source pressure magnitude, but also its spatial footprint. Because the absorbed energy follows the elongated plasma region, the induced pressure field inherits the same axial asymmetry. The thermoelastic source pressure therefore inherits a plasma-shaped spatial profile, not a spherical heat source distribution. This distinction determines the direction and spatial extent of the early mechanical loading that precedes bubble inception.

The gigapascal-scale source pressure greatly exceeds the ambient hydrostatic pressure and vapor pressure, indicating that the earliest liquid response is governed by a thermoelastic acoustic impulse rather than by an equilibrium phase-transition process. The acoustic relaxation of this compressive source field into tensile rarefaction loading and cavity inception is examined next.

### 3.4 Tensile loading and bubble inception

The pressure field in **Fig. 6** represents the local compressive source pressure generated by rapid plasma heating. Bubble inception, however, is governed by the subsequent dynamic pressure response following acoustic relaxation of the source field. **Fig. 7** shows the pressure trace obtained from the axisymmetric thermoelastic acoustic response model introduced in **Sec. 2.2** at the location of maximum source pressure, using the same time convention as Fig. 4, where $t=0$ denotes the pulse peak. The pressure first rises during the pre-peak portion of the pulse as plasma-mediated absorption increases the local temperature. After the pulse peak, the compressive field decays and transitions into a tensile excursion, creating the mechanical condition required for cavity inception.

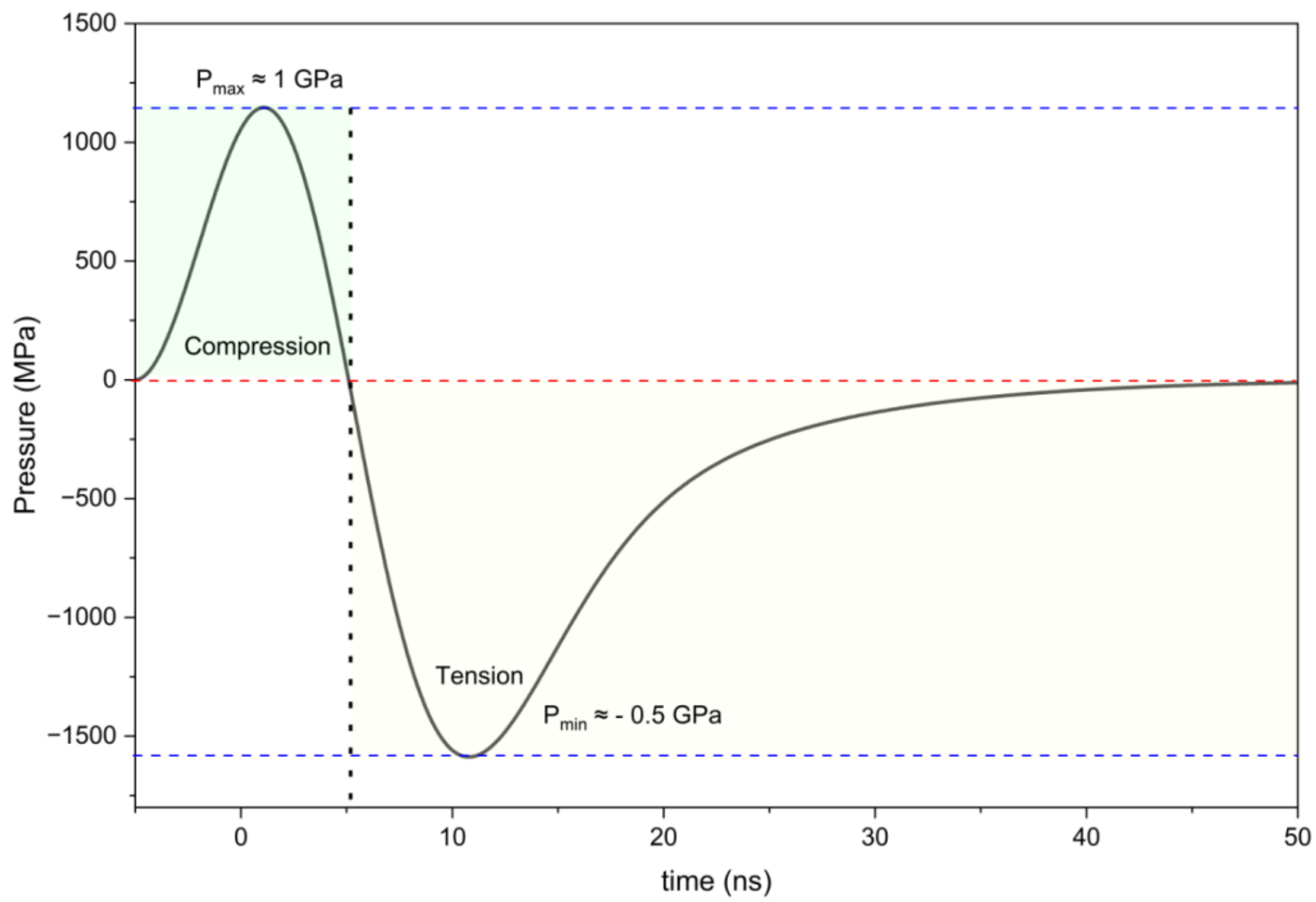

***Fig. 7. Compressive-to-tensile evolution of the dynamic thermoelastic pressure response.*** *The pressure trace is obtained from acoustic relaxation of the plasma-generated thermoelastic source pressure at the location of maximum source strength. The initially compressive response evolves into a tensile rarefaction phase; cavity inception occurs when the dynamic tensile pressure satisfies the mechanical cavitation criterion.*

This pressure evolution occurs under finite-duration stress-confinement conditions. For the nanosecond pulses and micrometer-scale heated volumes considered here, the acoustic relaxation time $\tau_{ac} = L/c_s$ is on the order of several nanoseconds, where $L$ is the characteristic size of the heated region and $c_s \approx 1500$m/s is the speed of sound in water. Because $\tau_{ac}$ is comparable to the laser pulse duration, the response should not be interpreted as an ideal instantaneous stress-confined pressure jump. In the present formulation, finite-pulse thermoelastic loading is represented by driving the acoustic response model with a Gaussian temporal source whose FWHM is equal to the laser pulse duration. The compressive-to-tensile transition occurs as the initially confined pressure relaxes through the emission of outward-propagating acoustic waves. As these waves leave the focal region, the local pressure at the center decreases and eventually enters a tensile regime. This behavior is consistent with thermoelastic stress generation in confined absorbers, where rapid local heating produces an initial compressive response followed by rarefaction during acoustic relaxation [37]. In the present model, cavity inception occurs when this dynamic tensile pressure satisfies the mechanical criterion in **Eq. (12).**

This result separates the physics of bubble inception from that of later bubble growth. The earliest cavity is initiated by a transient mechanically driven tensile pressure, rather than by an equilibrium thermal nucleation process. Thermal energy retained after plasma absorption can still contribute to subsequent expansion, but it is not required to trigger the initial cavity in the nanosecond breakdown cases considered here. Although the present model does not explicitly propagate the far-field shock wave, the early compressive stage should be interpreted as the near-field mechanical impulse associated with shock generation. The geometric consequence of this plasma-driven tensile loading is examined next.

### 3.5 Initial bubble geometry from plasma-mediated loading

In addition to initiating cavitation, plasma-mediated tensile loading determines the geometry of the inception region. **Fig. 8** presents the predicted initial bubble shapes for three representative laser energies, 2, 10, and 30 mJ. In all cases, the earliest cavity is elongated along the laser propagation direction rather than being spherical, forming a pear-like structure. This result isolates plasma-mediated energy deposition as the primary origin of bubble-shape anisotropy at inception.

This non-spherical geometry arises from the spatial transfer of asymmetry across successive stages of the model. Plasma shielding and moving breakdown produce an upstream-elongated absorption region, which is then transferred to the thermoelastic pressure field. The tensile phase of this pressure field defines the initial cavity boundary. As a result, the bubble does not begin from a spherical thermal nucleus, but from a mechanically loaded region already biased by plasma dynamics.

The degree of elongation increases with laser energy, reflecting stronger plasma localization and upstream energy deposition at higher energies. This trend is quantified by the aspect ratio, AR = width/length. The decrease in AR from 0.45 at 2 mJ to 0.25 at 30 mJ indicates increasing anisotropy with energy. The spherical radius used later for comparison with imaging-based measurements should therefore be interpreted as an equivalent radius, not as an assumption of spherical inception.

These results identify the plasma-mediated tensile field as the origin of the initial bubble geometry. Spherical symmetry becomes a useful later-stage approximation only after hydrodynamic expansion has partially smoothed the initial anisotropy. The relative contributions of the plasma-mediated, thermal-dominated, and coupled descriptions are compared in the following section.

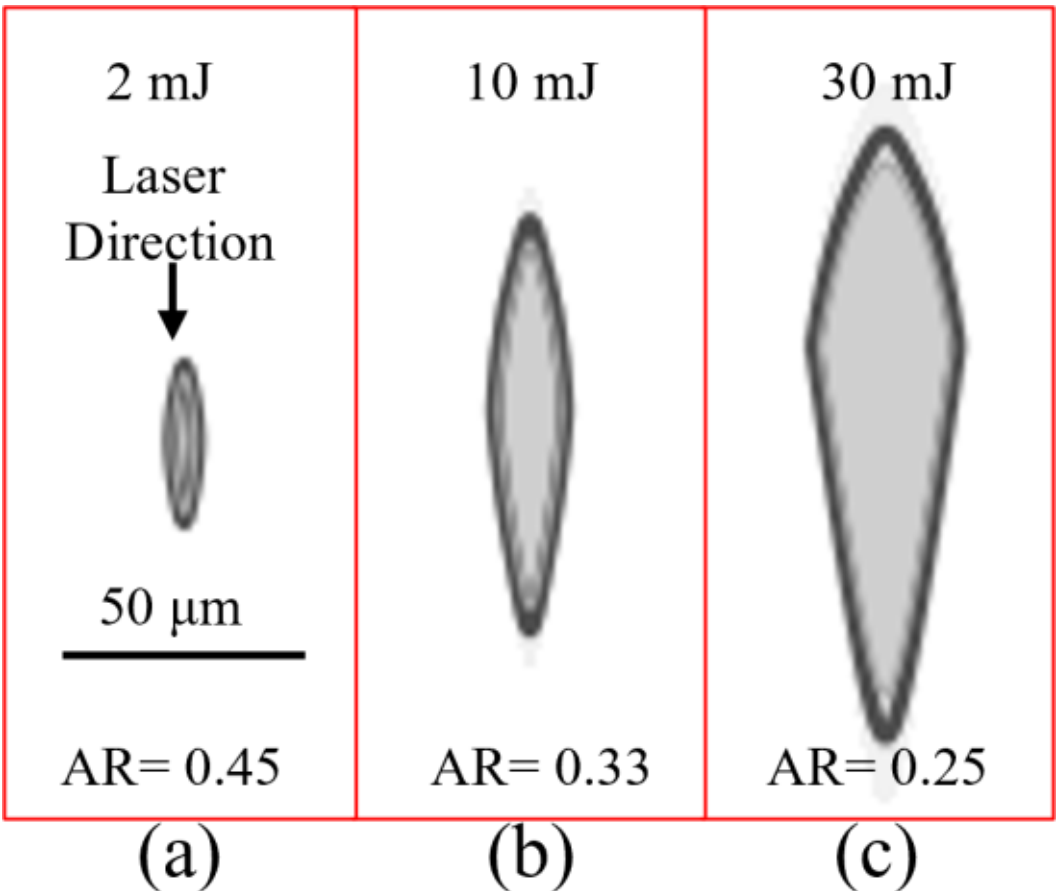

***Fig. 8. Predicted non-spherical inception geometry generated by plasma-mediated tensile loading.*** *Initial cavity shapes are shown for input laser energies of 2, 10, and 30 mJ. The cavity becomes increasingly elongated along the laser propagation direction as the energy increases, reflecting stronger shielding driven plasma localization. The aspect ratio is defined as* $AR = width/length$*; the equivalent spherical radius used later is a scalar comparison metric rather than an assumption of spherical inception.*

### 3.6 Initial bubble size and inception mechanism

The preceding sections identify a plasma-mediated pathway in which localized absorption produces thermoelastic loading, tensile relaxation, and a non-spherical inception region. The next question is whether this pathway also controls the earliest experimentally measurable bubble size. To test this, the initial bubble sizes predicted by the plasma-mediated, thermal-only, and coupled plasma–thermal models are compared with the experimental measurements of Jia et al. [4].

**Fig. 9** presents this comparison at t = 5 ns after the pulse peak, corresponding to the earliest experimentally resolvable frame in the reference dataset. It should be noted that the experimental bubble radii at this time point are obtained from literature-based extraction and fitting, and are therefore subject to uncertainties associated with image resolution, contour reconstruction, and the finite temporal resolution of the imaging system. This comparison is accordingly interpreted as an indicator of the dominant physical mechanism rather than a definitive quantitative benchmark.

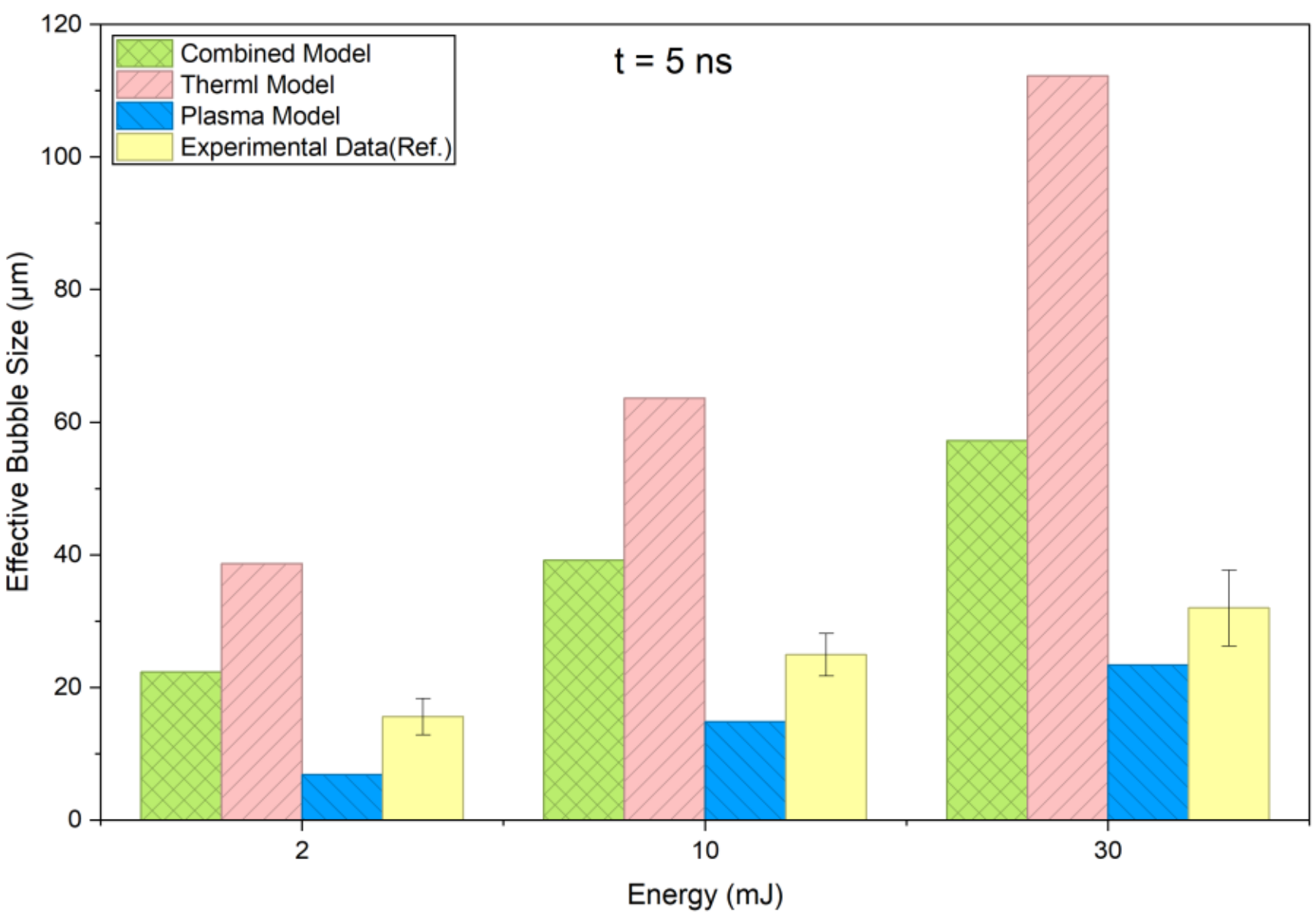

***Fig. 9. Initial bubble size predicted by the three formulations at the earliest resolved time.*** *Effective bubble radius at* $t = 5ns$ *is compared among the plasma-mediated, thermal-only, and coupled plasma-thermal models against experimental values extracted from Jia et al. The comparison evaluates which mechanism best captures the inception-scale bubble size rather than the later hydrodynamic growth.*

The thermal-only model significantly overpredicts the bubble size across all three energy levels (2, 10, and 30 mJ). This systematic overprediction has a clear physical origin: in the absence of plasma localization, the effective absorption coefficient $\alpha_{eff}$ distributes the absorbed energy over a volume substantially larger than the actual plasma region. The resulting temperature field is spatially diffuse, leading to superheating over a larger volume and an initial bubble size that substantially exceeds experimentally observed values. Because the total absorbed energy is the same in the thermal-only and plasma-mediated limiting cases, this systematic overprediction cannot be attributed to excessive total energy input. Instead, it reflects the spatially diffuse energy-deposition profile and the thermally activated inception criterion used in the thermal-only description. This overprediction is consistent with the fundamental limitation of purely thermal models: they cannot reproduce the sub-focal-volume concentration of energy deposition that is characteristic of plasma-mediated breakdown [1], [3].

In contrast, the plasma-mediated model yields the closest agreement with the experimentally measured bubble radii at moderate (10 mJ) and high (30 mJ) energies. The agreement improves with increasing energy, reflecting the fact that at higher energies, plasma formation and cascade ionization are more fully developed, and the thermoelastic mechanism dominates more decisively. At the lowest energy (2 mJ), the plasma-mediated model slightly underestimates the bubble size, suggesting that near the breakdown threshold, incomplete ionization reduces the thermoelastic impulse, and thermal contributions begin to play a supplementary role.

The coupled plasma-thermal model produces intermediate predictions that improve the agreement at low energy while maintaining reasonable accuracy at higher energies. This behavior is physically consistent: the coupled model captures both the localized mechanical impulse from plasma absorption and the additional contribution of residual thermal energy to early-stage vapor generation. At 2 mJ, where the plasma-mediated model alone underestimates the bubble size, the thermal augmentation provided by the coupled framework partially compensates for the weaker thermoelastic impulse. However, at higher energies, where the plasma mechanism dominates, the thermal contribution becomes secondary, and the coupled model converges toward the plasma-only prediction.

To quantify these differences, **Fig. 10** presents the root-mean-square error (RMSE) and normalized RMSE for each model. The RMSE is defined as

$$\mathrm{RMSE} = \sqrt{\frac{1}{N}\sum_{i=1}^{N}\left(R_{model,i} - R_{exp,i}\right)^2} \tag{34}$$

where $R_{model,i}$ and $R_{exp,i}$ denote the predicted and experimental bubble radii at the (i)-th energy level, respectively, and (N) is the total number of data points. In this study, (N = 3), corresponding to the three laser energies (2, 10, and 30 mJ) evaluated at t = 5 ns. Due to the limited number of data points, the RMSE values are interpreted as a comparative metric rather than a statistically converged measure.

To account for differences in magnitude across energy levels, the normalized RMSE is also reported, defined as

$$\mathrm{NRMSE} = \frac{\mathrm{RMSE}}{\overline{R_{exp}}} \tag{35}$$

where $\overline{R_{exp}}$ is the mean experimental bubble radius across all energy levels.

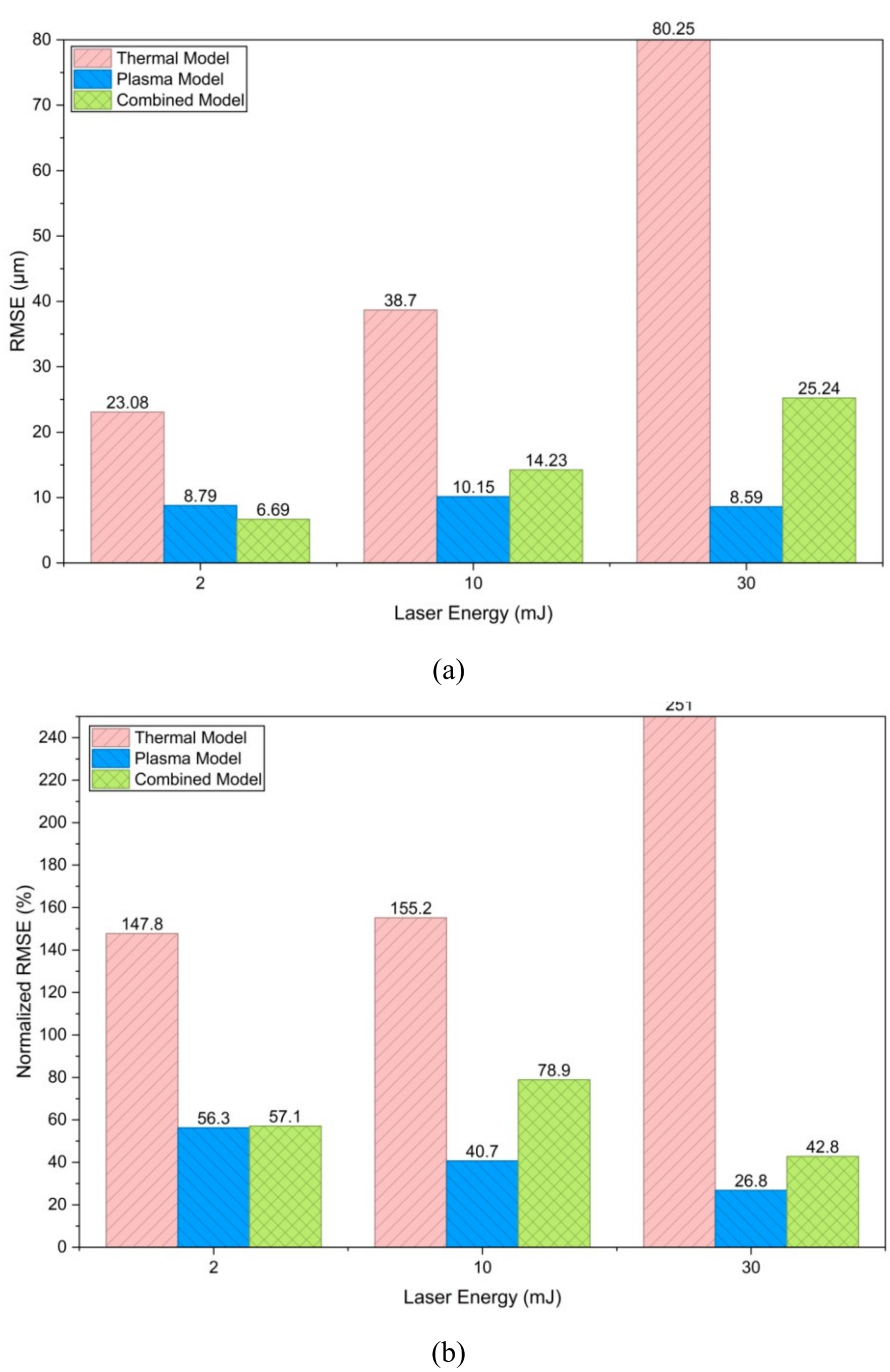

***Fig. 10. Error metrics for the initial bubble-size comparison.*** *(a) Root-mean-square error (RMSE) and (b) normalized RMSE (NRMSE) for the plasma-mediated, thermal-only, and coupled plasma-thermal models evaluated at t=5 ns across the three laser energies. Because only three energy levels are used, the metrics are interpreted as comparative indicators of model performance rather than statistically converged uncertainty estimates.*

The error analysis confirms the hierarchy suggested by the direct comparisons in **Fig. 9**. The thermal-only model exhibits the largest deviations, and the discrepancy increases strongly with laser energy. Specifically, the RMSE increases from 23.08 μm at 2 mJ to 80.25 μm at 30 mJ, while the NRMSE increases from 147.8% to 251%. This reflects the fundamental limitation of the thermal-only description under breakdown conditions. As laser energy increases, plasma shielding and moving breakdown concentrate the deposited energy into a localized and anisotropic region, whereas the thermal-only model distributes the absorbed energy via an effective absorption coefficient over a broader volume. The resulting heated region is therefore too diffuse, leading to an exaggerated initial bubble size.

In contrast, the plasma-mediated model shows substantially smaller error at the earliest resolved time. Its normalized error decreases with increasing laser energy, indicating that plasma formation, cascade ionization, and thermoelastic loading become more dominant as the breakdown becomes more fully developed. This trend confirms that the inception-scale bubble size is primarily controlled by the plasma-induced mechanical response rather than by purely thermal nucleation.

The coupled plasma–thermal model reduces the error relative to the thermal-only framework and improves the low-energy prediction, but it does not consistently outperform the plasma-mediated model at this single early time point. This observation does not diminish the value of the coupled framework. Instead, it reflects the physical reality that at t = 5 ns, the bubble remains within the mechanical inception regime, where the thermoelastic impulse remains the dominant driving force. The full advantage of the coupled model becomes apparent only when the subsequent thermally sustained growth is evaluated over the extended radius-time evolution in **Sec. 3.7**.

Together, **Figs. 9 and 10** support three conclusions regarding initial-stage bubble dynamics. First, under nanosecond breakdown conditions, bubble inception is primarily governed by plasma-induced thermoelastic loading. Second, purely thermal models systematically overestimate the initial bubble size because they do not reproduce the spatial confinement of plasma-mediated energy deposition. Third, thermal effects can contribute near the low-energy threshold condition, where the plasma impulse is weaker, but they remain secondary during the inception stage. The role of residual thermal energy in sustaining bubble growth beyond this initial period is examined next through the full temporal evolution of the bubble radius.

### 3.7 Long-term bubble dynamics: thermally sustained growth after plasma-driven inception

The preceding comparison shows that the earliest measurable bubble size is primarily controlled by plasma-induced thermoelastic loading, with thermal effects playing a secondary role near the low-energy threshold condition. At later times, however, the initial mechanical impulse has already propagated away from the focal region, while residual thermal energy remains in the liquid surrounding the nascent cavity. The long-time radius evolution therefore provides a separate test of whether the coupled framework can capture the transition from mechanically driven inception to thermally sustained growth. **Fig. 11** presents the temporal evolution of the effective bubble radius at 2, 10, and 30 mJ, compared against the experimental data of Jia et al. [4]. At the final comparison time of 2000 ns, the observed bubble radius increases from approximately 170 μm at 2 mJ to 260 μm at 10 mJ and 350 μm at 30 mJ, giving normalized radii of approximately 1.00, 1.53, and 2.06 when the 2 mJ case is used as the reference.

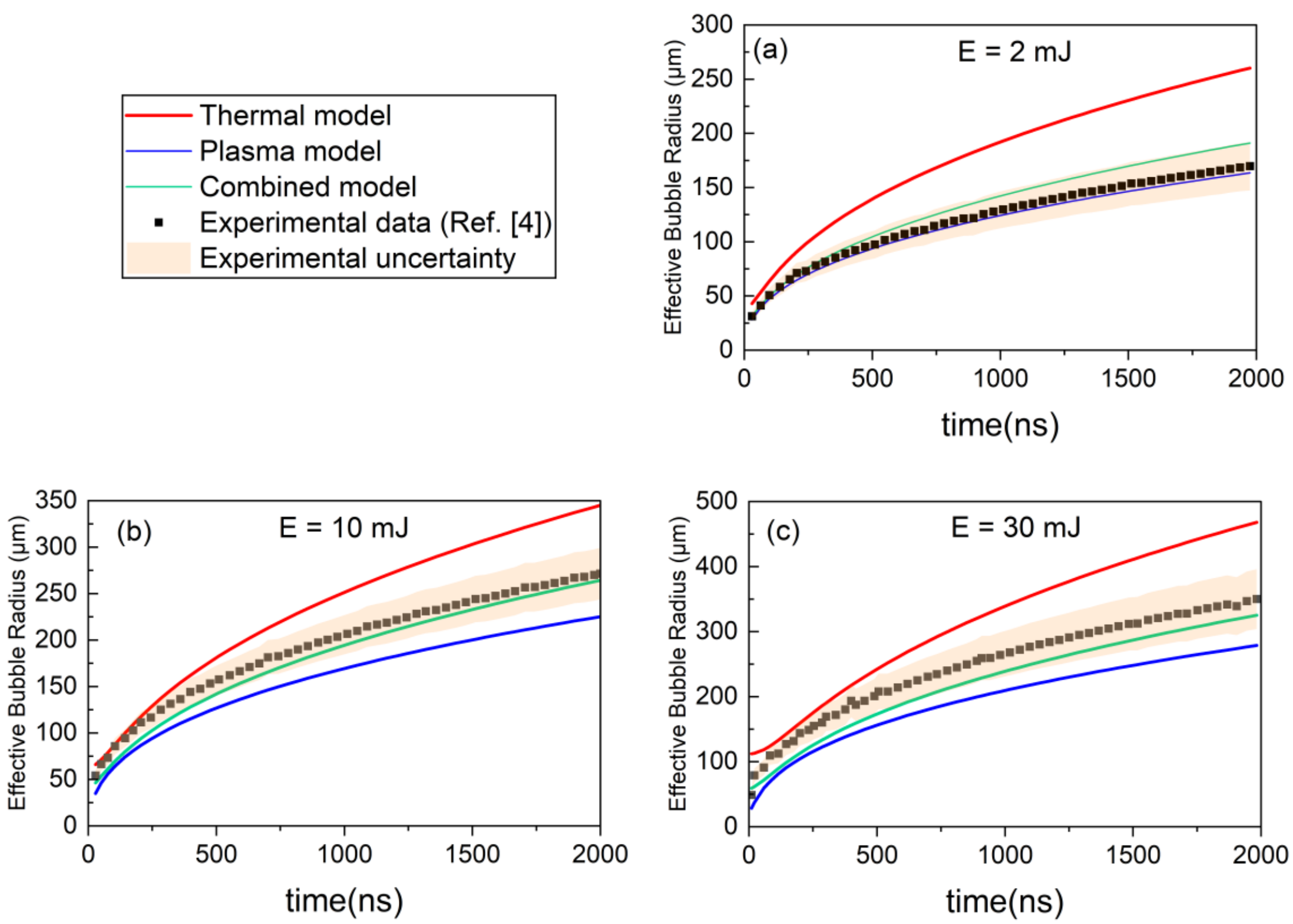

***Fig. 11. Temporal evolution of the effective bubble radius after inception.*** *Model predictions are compared with experimental radius–time data for laser energies of (a) 2 mJ, (b) 10 mJ, and (c) 30 mJ. The shaded regions represent experimental uncertainty. The coupled plasma–thermal model captures the transition from mechanically driven inception to thermally sustained growth more consistently than either limiting model.*

The plasma-mediated model reproduces the steep initial growth at all energies, but systematically underestimates the bubble radius at later times, with the deviation becoming more pronounced as the laser energy increases. This energy-dependent underprediction has a clear physical origin: higher laser energies deposit more residual thermal energy into the surrounding liquid through plasma absorption, while the plasma-mediated limiting case retains only the transient mechanical impulse. The neglected thermal reservoir therefore becomes increasingly important as input energy increases. This divergent behavior, in which the plasma-mediated model captures the initial expansion but increasingly underpredicts later growth, indicates that the mechanical impulse alone is insufficient to describe the full post-inception evolution.

The thermal-only model exhibits an opposite picture: its overprediction intensifies with energy over the full radius-time evolution. At later times, the overprediction grows from moderate at 2 mJ to severe at 30 mJ. This scaling arises because higher laser energies produce stronger, more spatially localized plasma absorption, concentrating the deposited energy into a smaller fraction of the focal volume. The thermal-only model, which distributes energy diffusely through $\alpha_{eff}$, creates an increasingly exaggerated mismatch with the actual energy distribution as the plasma develops more fully. The thermal-only model therefore performs worst where plasma formation is most vigorous, reinforcing the limitation of purely thermal descriptions of nanosecond breakdown.

The coupled plasma-thermal model provides the closest agreement across all energies and the full temporal range. At 2 mJ, where incomplete ionization produces a weaker thermoelastic impulse, the residual thermal contribution improves the predicted growth relative to the plasma-mediated limit. At 10 mJ, the model tracks the experimental data through both the steep early expansion and the gradual later growth, remaining within the experimental uncertainty band. At 30 mJ, the coupled model still outperforms both limiting descriptions, although the slight underestimation at the latest times may signal the onset of processes not captured in the present framework, such as non-equilibrium evaporation-condensation kinetics, evolving non-condensable gas content, or supercritical-state effects [23], [30], [38].

The physical basis for the coupled model's robustness is that the mechanical and thermal contributions originate from the same plasma-mediated energy deposition event rather than from two separately prescribed sources. During plasma absorption, inverse Bremsstrahlung generates a thermoelastic impulse that governs early expansion while also leaving residual heat that sustains later growth. As the pressure impulse decays, the retained heat continues to support vapor generation and bubble expansion, producing the gradual reduction in slope observed in the radius-time curves. This behavior is consistent with the 20 to 50 ns transitional phase identified experimentally by Jia et al. [4], in which the system evolves from moving breakdown dominated early dynamics toward thermal expansion.

In summary, the longtime radius evolution shows that the two limiting models fail in opposite directions: the plasma-mediated model underpredicts later growth because it lacks residual thermal support, whereas the thermal-only model overpredicts the radius because it lacks plasma localized energy deposition. The coupled framework captures the balance between these two effects, supporting the interpretation of nanosecond laser-induced cavitation as a plasma-seeded and thermally sustained process.

### 3.8 Physical interpretation and implications

Taken together, the present results support a broader interpretation of laser-induced cavitation under nanosecond breakdown conditions: bubble formation is neither a purely thermal phase-transition event nor merely a later hydrodynamic consequence of an already specified plasma source, but a coupled process in which breakdown morphology, thermoelastic loading, and residual heating jointly shape the birth and subsequent growth of the cavity. In this picture, plasma dynamics determine the spatial distribution of absorbed laser energy. This plasma-shaped energy deposition field defines both the magnitude and geometry of the thermoelastic source pressure, and the resulting mechanically driven cavity then evolves under the continuing influence of residual thermal energy.

This positioning is important in relation to existing studies. High-resolution experimental work has made increasingly clear that early cavitation can inherit strong spatial heterogeneity from the underlying breakdown process, while more detailed continuum and CFD-based models have shown that later bubble evolution may depend sensitively on phase change, gas content, and other thermodynamic processes once an initial bubble has been prescribed [4], [23], [24]. The present framework occupies the intermediate layer between these two ends of the problem by resolving how the initial bubble state emerges from nanosecond breakdown itself. The quantitative comparison in the present work is based on the time-resolved measurements of Jia et al. [4]; additional datasets spanning different focusing geometries, pulse durations, and liquid conditions will be needed to further test the transferability of the framework. Furthermore, the present post-inception model tracks the equivalent bubble radius rather than the full non-spherical interface evolution, which motivates future coupling with axisymmetric or three-dimensional CFD simulations. From this perspective, the framework complements both experimental diagnostics of breakdown-to-cavity transition and later-stage cavitation models by providing a mechanism-based bridge between them.

## 4. Conclusion

A coupled plasma-thermal framework has been developed for laser-induced cavitation in water under nanosecond optical breakdown conditions. By integrating free electron dynamics, inverse Bremsstrahlung absorption, thermoelastic acoustic response, residual thermal energy retention, and post-inception bubble dynamics, the model provides a physically consistent description of how localized optical energy deposition generates a thermoelastic acoustic response, triggers cavity inception through transient tensile rarefaction, and supports thermally sustained bubble growth.

The results show that the earliest cavity is primarily initiated by plasma-induced thermoelastic acoustic relaxation. Spatially confined plasma absorption generates a compressive source pressure whose subsequent rarefaction produces the transient tensile loading required for mechanical cavitation, while residual thermal energy sustains post-inception growth. The two limiting descriptions fail in opposite ways: the plasma-mediated model underpredicts later growth because it lacks residual thermal support, whereas the thermal-only model overpredicts inception size because, despite receiving the same total absorbed energy through absorbed energy matching, it distributes that energy over a broader heated volume. The coupled framework reconciles this contrast by treating the mechanical and thermal responses as staged outcomes of a common plasma-mediated energy deposition event.

The framework further establishes that the initial cavity inherits the anisotropic morphology of the plasma absorption region rather than emerging as a spherical nucleus, providing a mechanistic link between breakdown geometry and bubble birth state. The equivalent spherical radius used for Rayleigh-Plesset evolution is therefore a reduced order comparison metric rather than an assumption of spherical inception; full non-spherical interface dynamics remain a subject for future axisymmetric or three-dimensional CFD treatment.

These results support the interpretation of nanosecond laser-induced cavitation as a plasma-seeded and thermally sustained process. The model provides physically grounded initial conditions for multiscale modeling of laser driven material transport processes. Future extensions should incorporate non-equilibrium phase change kinetics, nonspherical interface evolution, and validation against additional experimental datasets spanning different focusing geometries and liquid conditions.

**Acknowledgements**

The authors would like to acknowledge the funding support from the National Science Foundation (NSF) under CAREER Award No. CBET-2441995, "Laser-Induced Bubble Dynamics and Jet Formation in Complex Fluids," from the CBET Fluid Dynamics Program.

**Conflict of Interest**

The authors declare no conflict of interest.